# Weak coupling of pseudoacoustic phonons and magnon dynamics in incommensurate spin ladder compound $Sr_{14}Cu_{24}O_{41}$

Xi Chen,[1] Dipanshu Bansal,[2] Sean Sullivan,[1] Douglas L. Abernathy,[3] Adam A. Aczel,[3] Jianshi Zhou,[1,4] Olivier Delaire,[5,2] Li Shi[1,4,*]

[1] Materials Science and Engineering Program, Texas Materials Institute, The University of Texas at Austin, Austin, Texas 78712, USA
[2] Materials Science and Technology Division, Oak Ridge National Laboratory, Oak Ridge, Tennessee 37831, USA.
[3] Quantum Condensed Matter Division, Oak Ridge National Laboratory, Oak Ridge, Tennessee 37831, USA.
[4] Department of Mechanical Engineering, The University of Texas at Austin, Austin, Texas 78712, USA.
[5] Department of Mechanical Engineering and Materials Science, Duke University, Durham, North Carolina 27708, USA.
*lishi@mail.utexas.edu

**Abstract**: Intriguing lattice dynamics has been predicted for aperiodic crystals that contain incommensurate substructures. Here we report inelastic neutron scattering measurements of phonon and magnon dispersions in $Sr_{14}Cu_{24}O_{41}$, which contains incommensurate one-dimensional (1D) chain and two-dimensional (2D) ladder substructures. Two distinct pseudoacoustic phonon modes, corresponding to the sliding motion of one sublattice against the other, are observed for atomic motions polarized along the incommensurate axis. In the long wavelength limit, it is found that the sliding mode shows a remarkably small energy gap of 1.7-1.9 meV, indicating very weak interactions between the two incommensurate sublattices. The measurements also reveal a gapped and steep linear magnon dispersion of the ladder sublattice. The high group velocity of this magnon branch and weak coupling with acoustic and pseudoacoustic phonons can explain the large magnon thermal conductivity in $Sr_{14}Cu_{24}O_{41}$ crystals. In addition, the magnon specific heat is determined from the measured total specific heat and phonon density of states, and exhibits a Schottky anomaly due to gapped magnon modes of the spin chains. These findings offer new insights into the phonon and magnon dynamics and thermal transport properties of incommensurate magnetic crystals that contain low-dimensional substructures.

I. **Introduction**

Incommensurate compounds consist of two or more interpenetrating sublattices with lattice periods incommensurate along one or more crystal axes. They belong to the broad class of aperiodic crystals,[1] which also includes the celebrated quasicrystals.[2] These compounds usually show anisotropies in electronic, optical, magnetic, and thermal properties. Examples include superconducting $Hg_{3-x}AsF_6$ containing two orthogonal and nonintersecting linear chains of mercury atoms,[3] intermetallic Nowotny chimney-ladder phases[4] with potential applications in optoelectronic[5] and thermoelectric[6-8] devices, and spin ladder compound $Sr_{14}Cu_{24}O_{41}$, where the $CuO_2$ spin chains and $Cu_2O_3$ two-leg spin ladders are incommensurate along the $c$ axis. This spin ladder compound exhibits superconductivity under high pressure when doped with Ca,[9] as well as a large thermal conductivity contribution ($\kappa_{mag}$) from magnons,[10] which are broadly defined as the energy quanta of collective excitations of the spin degrees of freedom.[11]

Because of its relevance to not only the thermal properties but also electronic, optical, and magnetic properties via coupling between different energy excitations,[11] the unusual lattice dynamics of incommensurate compounds have been investigated in a few theoretical studies over the past 40 years.[12-16] If the interaction between two incommensurate sublattices is sufficiently weak, the energy of the system remains invariant during rigid displacement of one sublattice relative to the other along the incommensurate axis.[13] Consequently, in the long wavelength limit, the energy cost vanishes for relative out-of-phase sliding motions of the two sublattices along the incommensurate direction. Therefore, instead of the three acoustic polarizations of conventional crystals, including one longitudinal acoustic (LA) and two transverse acoustic (TA) polarizations, such incommensurate substructures have been predicted to exhibit four acoustic polarizations.[13, 14] Two of them correspond to the atomic displacements of each of the two sublattices along the incommensurate direction, while the other two polarizations are conventional acoustic modes of the whole aperiodic structure with atomic motions polarized perpendicular to the incommensurate direction. The decoupled atomic displacement modes of the two sublattices become two LA and two TA polarizations when the modes propagate parallel and perpendicular to the incommensurate axis, respectively. Interactions between the two sublattices, including non-linear force constants as well as Coulomb forces between charged sublattices,[15] can result in a small energy gap for the out-of-phase sliding motion between the two sublattices, which thus becomes a very low-energy optical mode. In this case, there are still only three pure acoustic polarizations with zero-energy gap in the long wavelength limit, including two polarized perpendicular to the incommensurate axis, and the third acoustic polarization corresponding to the in-phase translation of both incommensurate structures along the incommensurate axis.

A few experimental studies have sought to test different theories of the unusual lattice dynamics in incommensurate structures.[17-23] Recently, infrared and Raman-active phonon modes with energy as low as 1-2 meV were observed by THz time-domain, Raman, and infrared (IR) spectroscopy measurements in $Sr_{14}Cu_{24}O_{41}$.[24] These low-energy Raman and IR-active modes were attributed to the small energy gap of the out-of-phase sliding motion of the two charged sublattices. However, the behaviors of these sliding modes away from the zone center have remained elusive, as previous inelastic neutron scattering (INS) studies[25-31] of $Sr_{14}Cu_{24}O_{41}$ have been mainly focused on magnetic excitations rather than lattice dynamics. Furthermore, existing experimental results on other incommensurate compounds were unable to resolve some of the predicted features of the lattice dynamics of incommensurate crystals, such as additional pseudo-acoustic modes or very low-lying optical modes polarized parallel to the incommensurate axis for wave vectors not only parallel to, but also perpendicular to the incommensurate axis.

Here, we report time-of-flight INS measurements of both phonon and magnon dispersions in $Sr_{14}Cu_{24}O_{41}$ single crystals. Our experimental results clearly reveal the presence of two distinct pseudo-acoustic or low-lying optical modes polarized along the incommensurate axis, when the phonon wave vector is either parallel or perpendicular to the incommensurate axis. The small energy gap for the sliding mode is resolved and quantified. Moreover, our INS measurements reveal high-energy optical phonon modes with nearly linear dispersion and high group velocities, in addition to the extremely steep magnon dispersions of the spin-ladders, already reported in prior INS studies.[25, 30, 31] In addition, we evaluate the magnon specific heat below 300 K based on the measured total specific heat and phonon density of states (DOS). These findings shed light into the unique phonon and magnon dynamics of this incommensurate spin-ladder compound, and can lead to further insights on the thermal transport properties of incommensurate crystals in general.

## II. Experimental Methods

$Sr_{14}Cu_{24}O_{41}$ single crystals were grown with a traveling solvent floating zone method.[32] The crystal structure was verified via X-ray diffraction (XRD) measurements of a pulverized sample. The orientation and quality of crystals were examined by back-reflection Laue XRD. The low-temperature thermal conductivity ($\kappa$) was measured with a steady-state method. The temperature dependent specific heat was measured using a Physical Properties Measurement System (Quantum Design) in the temperature ($T$) interval between 1.9 and 300 K.

INS measurements were conducted on both crystals and a powder sample, with the time-of-flight Wide Angular-Range Chopper Spectrometer (ARCS) at the Spallation Neutron Source at Oak Ridge National Laboratory (ORNL).[33] Two $Sr_{14}Cu_{24}O_{41}$ single crystals were co-aligned to achieve a total mass of approximately 6 g. The mosaicity of the co-alignment was about 1 degree. Several incident neutron energies, $E_i$ = 30, 50, 150 meV, were used to cover the wide dynamic range of low-energy pseudo-acoustic modes as well as high-energy optical phonons and magnons. The crystals were mounted inside a closed-cycle helium refrigerator with the (0*KL*) crystallographic plane horizontal. The crystal measurements were performed at $T$ = 5 K and 140 K. In addition, the energy gap of the sliding mode was measured using cold (CG-4C) triple-axis spectrometer at High Flux Isotope Reactor, ORNL. The CG-4C measurements were performed using the PG002 monochromator and analyzer, with a constant final energy $E_f$ = 5.0 meV, and collimation settings of 270'–270'–80'–270' at (011) Bragg peak corresponding to the chain sublattice.

In addition, the phonon DOS was measured on a powder sample at 5 K, 70 K, 140 K, 210 K and 300 K. Two incident neutron energies, $E_i$ = 50 and 200 meV, were used at each temperature to achieve a higher resolution at the lower energy and to probe the full phonon spectrum, respectively.

The data from both single crystal and powder measurements were normalized by the total incident flux and corrected for detector efficiency. Subsequently, the data were transformed from instrument coordinates, including the 2-theta and time of event, to the physical momentum transfer, **Q**, and energy transfer, $E$, using the MANTID reduction software.[34] For the powder sample, the scattering from the empty container was processed identically and subtracted from the sample measurements. The experimental phonon DOS was analyzed within the incoherent scattering approximation, and corrected for multi-phonon scattering effects.[35] The data near the elastic peak were removed in order to eliminate strong elastic signals, and a Debye-like quadratic energy dependence was used to extrapolate the phonon DOS at low energy $E$ < 3 meV for $E_i$ = 50 meV.

Two-dimensional intensity maps for the dynamical structure factor, $S(\mathbf{Q},E)$, were obtained from slices of the four-dimensional dataset using the HORACE software.[36] In order to improve the appearance of these maps, a Gaussian smoothing filter was applied to the numerical data, with a standard deviation of 2 pixels and a filter size of $2 \times 2$ pixels. Miller indices are henceforth expressed as $[HKL]_C$ for the chain sublattice and $[HKL]_L$ for the ladder sublattice.

### III. Results and Discussions
#### A. Crystal structure and neutron diffraction of $Sr_{14}Cu_{24}O_{41}$

The unit cell of $Sr_{14}Cu_{24}O_{41}$ contains two incommensurate sublattices along the $c$ axis, which are $CuO_2$ chains and $Cu_2O_3$ two-leg ladders, as illustrated in Fig. 1(a). Planes of $CuO_2$ chains are stacked alternately with planes of $Cu_2O_3$ ladders. One-dimensional Sr chains are present between the $Cu_2O_3$ ladders and $CuO_2$ chains, and are coordinated and commensurate with the ladder sublattice to form $Sr_2Cu_2O_3$. The lattice parameters are $a$ = 11.469 Å, $b$ = 13.368 Å, and $c_L$ = 3.931 Å for the ladders and $c_C$ = 2.749 Å for the chains. Thus, the $c$ axis is incommensurate for the ladder and chain sublattices,[37] with a ratio of approximately $\alpha = \sqrt{2} : 1$. Due to the incommensurate crystal structure, two sets of Bragg peaks associated with the chain and ladder sublattices, respectively, are observed in the $(0KL)$ plane in the elastic scattering channel, as shown in Fig. 1(b). The ladders contribute strong peaks at $(0K\text{-}2)_L$ in their corresponding reciprocal lattice. In comparison, relatively weak diffraction peaks from the chains are observed at $(0K\text{-}1)_C$ and $(0K\text{-}2)_C$, equivalent to $(0K\text{-}1.43)_L$ and $(0K\text{-}2.85)_L$. Similar diffraction patterns were also observed in electron diffraction measurements on $Sr_{14}Cu_{24}O_{41}$.[38] The obtained neutron diffraction pattern is consistent with the calculated results, as shown in Figs. 1(c,d). It should be noted that the reciprocal lattice points marked by the cricles and squares in Figs. 1(c,d) are the zone centers. On the other hand, the crosses represent the zone boundaries. The thermal conductivity of a $Sr_{14}Cu_{24}O_{41}$ single crystal in this work agrees well with the results obtained by Hess *et al.*,[39] as discussed in Appendix A.

### B. Phonon dispersion of $Sr_{14}Cu_{24}O_{41}$

Figure 2(a) shows the $S(\mathbf{Q},E)$ along $\mathbf{q} = [00L]$. For a net momentum transfer from the incident neutron of $\mathbf{Q} = \mathbf{G}_{HKL}+\mathbf{q}$, the measured neutron-phonon scattering intensity increases with $|\mathbf{Q} \cdot \boldsymbol{\varepsilon}_i^\alpha|^2$, where $\boldsymbol{\varepsilon}_i^\alpha$ is the polarization vector of the $i^{th}$ atom in the unit cell for a particular phonon mode $\alpha$.[40] Consequently, the phonon modes with strong intensity in the Brillouin zone around $\mathbf{G}_{HKL} = (00L)$ are the longitudinal modes with $\mathbf{q}$ along $[00L]$, or the transverse modes polarized along $[00L]$ and propagating normal to it. It should be noted that the phonon intensity generally increases as $|\mathbf{Q}|^2$, while the magnon intensity vanishes at large $|\mathbf{Q}|$ because of the magnetic form factor.[41] As can be seen in Fig. 2(a), two separate LA-like phonon branches are clearly observed. One LA-like branch extends up to about 20 meV with a minimum energy at $(00-2)_L$ or $(00-4)_L$, indicating that this mode, referred as p-$LA_L$ hereafter, originates from the atomic motion of the ladders polarized along the incommensurate direction. The other LA-like branch extends up to about 30 meV and reaches a minimum energy at $(00-2.85)_L$ or $(00-2)_C$, so that this mode (p-$LA_C$) is associated with the atomic motion of the chains. The group velocities of the very linear LA-like modes are $v_C = 7300$ m s$^{-1}$ for the chains and $v_L = 4000$ m s$^{-1}$ for the ladders, as listed in Table I. The velocities of the acoustic-like modes depend on the mass density ($\rho$) and the elastic constant ($K$) of the two sublattices as $v = (K/\rho)^{1/2}$.[21] With a mass density $\rho_C = 1.51$ g cm$^{-3}$ for the chains and $\rho_L = 3.86$ g cm$^{-3}$ for the ladders, the ratio of elastic constants between chain ($K_C$) and ladder ($K_L$) sublattices is thus estimated to be $K_C : K_L = 1.24:1$. The larger elastic constant for the chain sublattice is consistent with the smaller 2.749 Å distance between the nearest Cu atoms along the $c$ axis for the chains as compared to the corresponding value of 3.931 Å for the ladders. In the low-frequency limit, the sound velocity of the pure LA mode associated with the global translational symmetry is given as

$$v_s = \sqrt{(m_C v_C^2 + m_L v_L^2)/(m_C + m_L)}, \quad (1)$$

where $m$ is the unit-cell mass of either sublattice as denoted by the $C$ or $L$ subscript.[41] The obtained value is about 5200 m s$^{-1}$ according to Eq. (1). In comparison, the longitudinal sound velocity extracted from the longitudinal elastic constant measured by an ultrasonic technique is about 6400 m s$^{-1}$,[42] which falls between the two group velocity values observed in INS measurements. It is noted that this extracted value is slightly higher than the above calculated value of the LA mode for the global translation of both sublattices. Since Eq. (1) is a simplified model for calculating the pure LA mode of the whole aperiodic structure, the calculation result can be somewhat different from the measurement result of the actual sample structure, which can deviate slightly from the structure considered in the theoretical calculation. In addition, the $c$-polarized, pure LA modes of the whole aperiodic structure are observable in theory only in the (000) zone of the aperiodic structure. However, the negligible inelastic scattering intensity compared to the high elastic background, limited detector coverage, and kinematic constraints of INS measurements at the (000) zone center make it difficult to observe the pure LA phonons in the INS data and measure their group velocity.

A linear extrapolation of the two LA-like branches down to the (00$L$) zone center indicates a non-zero energy gap [Fig. 2(b)], although the energy gap is partially obscured in Fig. 2(a) because of the strong Bragg peaks for the (00$L$) zones. To measure the energy gap directly, constant-**Q** scans at the (011)$_C$ zone center were carried out with the triple-axis spectrometer. As shown in Figs. 2(c) and (d), the energy gap is found to be about 1.7 meV at 300 K, and increases slightly with the decrease of temperature. This energy represents the cost of translating one sublattice against another rigid sublattice in the long wavelength limit, and is thus expected to be the same for sliding either the chain or the ladder sublattice relative to the other sublattice, which correspond to the two observed different LA-like, but essentially low-lying optical modes in Fig. 2(a).

Long-range Coulomb interaction between the two ionic sublattices can result in the observed energy gap in the sliding mode of incommensurate compounds. The gap value ($\omega_{gap}$) can be calculated as

$$\omega_{gap} = \sqrt{\frac{4\pi n_C q_C^2}{m_C}\left(1 + \frac{c_L}{c_C}\frac{m_C}{m_L}\right)}, \qquad (2)$$

where $n$ and $q$ are the atomic mass density and charge of each sublattice as denoted by the C and L subscripts.[15] This equation has been used to calculate a gap value of 3.7 meV for $Sr_{14}Cu_{24}O_{41}$ in a prior work.[24] The energy gap obtained from the INS measurements lies in between the calculated value and the energies of the Raman and infrared active modes observed in $Sr_{14}Cu_{24}O_4$,[24] and is larger than the value of 0.1 meV observed in $Hg_{3-x}AsF_6$.[43]

Figure 3 shows the dispersion of the two low-lying optical modes in the (*H*0*L*) plane. The chain excitation is much less dispersive along *a* than those along *c* because of relatively weak inter-chain interaction compared to intra-chain interaction. The much larger group velocities along *c* than along *a* can explain the highly anisotropic lattice thermal conductivity for $Sr_{14}Cu_{24}O_{41}$ below 40 K, where phonons dominate the contribution to $\kappa$.

The earlier INS experimental observation of additional acoustic-like modes in $Hg_{3-x}AsF_6$ stimulated the initial theoretical studies of lattice dynamics of incommensurate structures.[13, 17, 43] The theoretical studies have suggested experimental searches of new acoustic-like modes in Nowotny chimney-ladder structures such as higher manganese silicides (HMS). Recent INS measurements of HMS have indeed discovered the presence of a very low-lying optical polarization associated with the twisting motion of the Si ladder sublattice relative to the Mn chimney sublattice.[40] The twisting mode is observed near the Bragg peaks of the Si sublattices, with a component of the atomic motion polarized along the incommensurate direction.

Besides $Hg_{3-x}AsF_6$ and HMS, the appearance of additional acoustic-like branches along the incommensurate crystal axis has been observed in other incommensurate host-guest crystals including Rb-IV,[22] *n*-nonadecane-urea,[23] [Pt(*en*)$_2$][Pt(*en*)$_2$I$_2$](ClO$_4$)$_4$,[44] and layered superconducting cuprates Bi$_2$Sr$_2$CaCu$_2$O$_{8+\delta}$ and Bi$_2$Sr$_2$CuO$_{6+\delta}$.[20, 21] However, the measurement data for these incommensurate structures have not verified the presence of two decoupled TA-like modes polarized along the incommensurate axis.

In comparison, the INS data in Fig. 4(a) essentially show a TA-like mode of the ladders for polarizations along *c* and wave vectors along *a*. However, this combination of atomic polarization and wave vectors for the chains yields the dispersionless INS data in Fig. 3, which resemble the typical behavior of optical phonons instead of acoustic phonons. In addition to these TA-like or low-lying transverse optical modes polarized along *c*, the INS measurements along [*H* 16 0] reveal the pure TA mode polarized along *b* for wave vectors along *a* [Fig. 4(b)], and low-energy optical phonon modes in the range of 7-15 meV along [*H* 16 -2]$_L$ [Fig. 4(c)].

To search for clear experimental support of the theoretical predictions[12, 13] of two TA-like branches polarized along the incommensurate axis, we investigate $S(\mathbf{Q},E)$ for phonons propagating along *b*. The linear pure LA branch near (0 16 0) extends to about 10 meV at the zone boundaries at (0 15 0) and (0 17 0), as shown in Fig. 5(a). For INS data near a (00*L*) zone with large *L*, the data along *K* contain mainly transverse polarizations along *c*. As shown in Fig. 5(b), a TA-like branch near (002)$_L$ extends to ~7 meV. This branch is the *c*-polarized TA-like branch associated with the ladders. As the INS data in Fig. 5(c) are plotted with $L_C$= -2, the low-energy branch up to ~4 meV is the TA-like or very low-lying optical branch associated with the chains. It is noted that the *c*-polarized TA-like mode of the ladders exhibits a larger group velocity than that of the chains as listed in Table I, suggesting stiffer force constants between the ladders along *b* than those between the chains along the same direction. Nevertheless, the *c*-polarized low-lying optical mode of the chains is more dispersive or TA-like for wave vectors along the *b* axis than along the *a* axis. This observation is consistent with the weaker edge-to-edge interaction between the chains along the *a* axis than their face-to-face interaction along the *b* axis.

The observation of two TA-like, very low lying optical modes polarized along the incommensurate axis is striking, but consistent with the theoretical model of Axe and Bak[13]. In $Sr_{14}Cu_{24}O_{41}$, *c*-polarized atomic motion in one ladder causes more energy perturbation in the adjacent commensurate ladders than in the nearby incommensurate chains. Consequently, two different TA-like or very low-lying optical modes arising from dominantly the atomic motion of one of the two sublattices can be observed in the INS data around the elastic peaks of the corresponding sublattice. While this feature was not reported in prior studies, the current measurement results show the presence of a more dispersive *c*-polarized TA-like mode of the ladders, as well as a less dispersive, but still TA-like, very low-lying *c*-polarized optical mode of the chains, when the wave vector is perpendicular to the incommensurate axis.

In addition to the two acoustic-like or very low-lying optical modes polarized along the *c* axis for wave vectors both parallel and perpendicular to *c*, which are observed near the zone centers, some *c*-polarized low-lying optical phonons with energy gaps less than 5 meV have been observed at the zone boundary, as described in Appendix B.

Furthermore, the INS data show additional unique features for phonons polarized perpendicular to *c*. An acoustic mode is observed near the (0 16 0) zone with **q** along [00*L*], as displayed in Fig. 6(a). This mode is a TA phonon polarized along *b* with **q** along *c*, and shows a group velocity of 2100 m s$^{-1}$. Some optical phonons ranging from about 7 to 70 meV are shown in Figs. 6(b-d). A relatively low-lying optical polarization [Fig. 6(b)] exhibits a "W" shape, which is centered at (07-2)$_L$. One optical mode near (04-2)$_L$ appears in the energy range between 16 and 22 meV [Fig. 6(c)]. A unique feature of this optical branch is its nearly linear dispersion, which is equivalent to a relatively high group velocity of 1800 m s$^{-1}$. Another high-energy optical mode shows a maximum energy of ~70 meV at $L_L = 3$ or 5, and bends down to 56 meV at the zone boundary with a high group velocity of 2800 m s$^{-1}$ [Fig. 7(d)]. The high-energy optical mode should be due to the oxygen vibrations bounded to Cu because oxygen is the lightest ion and Cu-O is the shortest bond in the structure.[45] The shape of this mode is similar to the longitudinal oxygen bond-stretching mode in the $CuO_2$ plane observed in doped $La_2CuO_4$ compounds.[46] In addition, two Raman-active modes at 68 and 72 meV have been found in $Sr_{14}Cu_{24}O_{41}$,[47] which are assigned as breathing modes of oxygen atoms in the ladder, and vibrations of oxygen in the chain along the *c* axis, respectively. Since the Sr array is commensurate with the ladder sublattice, it is expected that Sr atoms are involved in the observed acoustic and optical modes near the Bragg peaks of the ladders.

### C. Magnon dispersion of $Sr_{14}Cu_{24}O_{41}$

The magnetic interactions are different in the two sublattices of $Sr_{14}Cu_{24}O_{41}$. In $Cu_2O_3$ ladders, the Cu $S=1/2$ spins are strongly coupled via the 180° Cu-O-Cu superexchange, while the magnetic interactions are weak in the array of $CuO_2$ $S = ½$ spin chains due to the 90° Cu-O-Cu configuration. The magnon dispersion of $Sr_{14}Cu_{24}O_{41}$ along $c$ is shown in Fig. 7(a). The stiff dispersions at $L_L = -(0.5+n)$, where $n$ is an integer number, show a large energy gap of ~31 meV. These dispersions are associated with spin-triplet excitations of the ladders.[25] Even-leg ladders have spin-liquid ground states due to their purely short-range spin correlation.[48] A spin gap is formed since a finite energy is needed to create an $S = 1$ spin excitation in the ladder.[49] These excitations are particularly clearly observed at low energy, as shown in Fig. 7(b). While the magnon gap has been determined from previous INS studies,[25, 30, 31] our INS measurements have mapped out about half of the dispersion and have shown the large bandwidth of these spin triplet excitations. In addition, the magnon dispersions of the chains are nearly dispersionless and show energy gaps of about 9 and 11 meV, as discussed in Appendix C.

The steep dispersion of the ladders indicates the large group velocity for magnons in the spin ladders. In addition, the INS data of Fig. 7(a) show that there is little overlap between acoustic-like branches and ladder magnons. On the other hand, the high energy optical modes of the ladders cross the magnon dispersion at about 55 and 72 meV as shown in Fig. 7(b). Since the magnon-phonon interaction is usually achieved via a two-magnon, one-phonon scattering process, the magnon-acoustic phonon scattering is unlikely to occur in $Sr_{14}Cu_{24}O_{41}$ due to the small energy overlap. Instead, magnon-optical phonon scattering is expected to be the main scattering mechanism at sufficiently high temperatures where the ladder magnons are populated.

### D. Phonon and Magnon specific heat contributions of $Sr_{14}Cu_{24}O_{41}$

While the observed magnon dispersion is in general agreement with prior reports, the magnon contribution to the specific heat has been determined from the measured phonon density of states and total specific heat in this work. Since the scattering intensity from magnons is limited to low $|Q|$,[41] we used high $|Q|$ data between 3 and 6.5 $Å^{-1}$ for 50 meV neutron energy and between 7 and 12 $Å^{-1}$ for 200 meV energy to obtain the phonon DOS. Figure 8(a,b) show the temperature dependence of neutron weighted phonon DOS with $E_i$ = 50 meV and 200 meV, respectively. We observe two phonon peaks with a low energy peak at ~30 meV and a high energy peak at 70 meV. The phonon cutoff energy is about 100 meV, in agreement with INS phonon dispersion results. With the increase of temperature, a slight softening of phonon is observed. The obtained phonon DOS was used to calculate the phonon specific heat ($C_p$) of $Sr_{14}Cu_{24}O_{41}$, as discussed in Appendix D. As shown in Fig. 8(c), a rapid upturn below ~10 K has been observed in the $C_{p,Exp}$ but not in the $C_{p,Calc}$ in the $C_p(T)/T^3$ plot, mainly because of the quadratic fitting used to obtain the phonon DOS at energies lower than 3 meV. Such a feature has also been observed in other incommensurate modulated compounds, which originates from the gapped low-energy phonon modes of the incommensurate structure.[50] Above 50 K, the discrepancy between $C_{p,Exp}$ and $C_{p,Calc}$ is mainly due to the contribution of magnons to the total measured $C_{p,Exp}$. Figure 8(d) shows the difference between the measured $C_p$ and calculated $C_p$. A low energy peak at about 100 K is observed, akin to a Schottky contribution from the gapped magnon dispersions of the chains. This peak temperature is comparable to the low-energy magnon gap, about 90 K. The magnon contribution is about 0.03 J g$^{-1}$ K$^{-1}$ at 300 K, about 7% of the total specific heat, which is of the same order of magnitude of the calculated $C_p$ of spin ladder compound Ca$_9$La$_5$Cu$_{24}$O$_{41}$.[51] At a temperature of 100 K, the magnon specific heat is found to be about 15% of the total $C_p$.

## IV.  Conclusions

INS measurements of the spin ladder compound Sr$_{14}$Cu$_{24}$O$_{41}$ clearly show the presence of two weakly coupled LA-like modes, emerging from the ladder and chain sublattices, respectively, along the incommensurate $c$ axis. In the long wavelength limit, the sliding mode exhibits a small energy gap of 1.7-1.9 meV due to the weak interaction between the two sublattices. Moreover, our measurements also reveal two decoupled $c$-polarized TA-like or very low-lying optical modes associated with the two sublattices for wave vectors perpendicular to the incommensurate axis. This latter feature provides further experimental support of prior theoretical predictions of the lattice dynamics of incommensurate compounds. A number of optical phonons are present in the energy ranging from 5 to 70 meV, some of which show relatively high group velocities. In addition, the steep magnon dispersion of the ladder sublattice has been mapped out from 31 to 120 meV. The high group velocity of these ladder-derived magnons and their weak coupling with acoustic phonons are responsible for the high $\kappa_{mag}$ in Sr$_{14}$Cu$_{24}$O$_{41}$. The magnon contribution to the specific heat has been evaluated from the measured $C_p$ and phonon DOS. A Schottky anomaly has been found in the obtained magnon specific heat, due to the gapped magnon dispersions of the chains. At a temperature of 100 K, magnons can contribute to ~15% of the total specific heat. These findings provide new insights into the phonon and magnon dynamics in incommensurate compounds, and may enable further theoretical analysis to explain these observations. Moreover, two-dimensional (2D) van der Waals heterostructures,[52] which are being actively investigated, are essentially incommensurate crystals, and can potentially exhibit similar phonon dynamics features as those observed here for the spin ladder compounds, which consist of weakly coupled low-dimensional incommensurate substructures.

## Acknowledgements

This work was supported by US Army Research Office (ARO) MURI award W911NF-14-1-0016. Neutron scattering measurements and analysis (D.B. and O.D.) were supported by the U.S. Department of Energy, Office of Science, Basic Energy Sciences, Materials Sciences and Engineering Division, through the Office of Science Early Career Research Program of O. D. (DE-SC0016166). The use of Oak Ridge National Laboratory's Spallation Neutron Source was sponsored by the Scientific User Facilities Division, Office of Basic Energy Sciences, U.S. Department of Energy. The authors thank helpful discussion with David G. Cahill and Yaroslav Tserkovnyak.

## APPENDIX A: THERMAL CONDUCTIVITY OF SR14CU24O41 SINGLE CRYSTAL

The thermal conductivity measurement results of $Sr_{14}Cu_{24}O_{41}$ single crystals reported by Hess *et al.*[39] and those obtained in the present work are shown in Fig. 9. The two measurements are in good agreement and show a striking anisotropy in both the absolute value and temperature dependence of $\kappa$. The $\kappa$ along $c$ shows a broad peak at about 140

K in addition to the phonon peak at a lower temperature. This feature has been attributed to the fast-propagating magnetic excitations along the spin ladders.[10,39] Below 40 K, the main contribution to $\kappa$ comes from phonons due to the presence of relatively large magnon gaps and a high electrical resistivity.[25, 53] The lattice thermal conductivity is found to be much larger along $c$.

**APPENDIX B: LOW-LYING OPTICAL PHONONS AT THE ZONE BOUNDARY**

Figure 10(a) shows INS data obtained near the forbidden Bragg peaks at the (01$L$) zones. The data arise from $c$-polarized, low-lying optical modes at the Brillouin zone boundary. The energy gaps of the optical modes are further confirmed by the constant-**Q** scan data measured at **Q** = (01-2)$_L$ and **Q** = (01-2)$_C$, as shown in Fig. 10(b). The peaks in the Fig. 10(b) are fitted to obtain the energy gaps, which are below 5 meV. Figure 10(c) and (d) show the INS data obtained along [$H$1-2]$_C$ and [$H$1-2]$_L$. Very flat INS intensity is observed along [$H$1-2]$_C$ in the range of -1.5 < $H$ < 1.5 with $E \approx$ 3 meV. In comparison, the phonon dispersion of the ladders along [$H$1-2]$_L$ is clearly evident.

**APPENDIX C: MAGNON DISPERIONS OF THE CHAIN SUBLATTICE**

In previous INS studies of $Sr_{14}Cu_{24}O_{41}$, the spin coupling energies for the ladders were found to be $J_{\parallel}$ = 130 meV along the ladder and $J_{\perp}$ = 72 meV along the rung.[25] The dominant intra-dimer exchange of the chains was evaluated to be $J \approx$ 10 meV, and the inter-chain exchanges were found to be $J_{\parallel} \approx -1.1$ meV and $J_{\perp} \approx 1.7$ meV.[28] The microscopic origin of dimerization is suggested to be the localization of holes at low temperature on structurally modulated chains.[54] The formed super-structure on the chains has a periodicity of $5c_C$, which corresponds to dimers of two spins coupled across one hole with neighboring dimers separated by two holes.[29,55]

Figure 11 shows the low-energy magnon dispersions of the chains, which are weakly dispersive along the *a* and *c* axes. These low-energy magnons have also been observed in previous INS measurements.[25, 27-30] As shown in Fig. 11(a), two different dispersions have been observed, which show energy gap values of about 9 and 11 meV, respectively. These dispersions have a periodicity of $0.2 \times (2\pi/c_C)$, suggesting that they originate from the aforementioned dimerized super-structure of the chain sublattice. Our results are in agreement with the INS data reported by Matsuda *et al*.[29] The energy gap of this magnon dispersion results from a dimerized state in the antiferromagnetic chains as suggested by both theoretical[56] and experimental studies.[29] Figure 11 (b,c) show the magnon dispersions along $[H\ 1\ 0.7]_C$ and $[H\ 1\ 0.2]_C$, respectively. One branch is shifted by one reciprocal lattice unit $2\pi/a$ with respective to the other. These two magnon branches result from the spin interaction of the chains along *a*.[28, 29]

**APPENDIX D: CALCUATION OF SPECIFIC HEAT FROM PHONON DOS**

The phonon DOS measured at 5 K was used to calculate the phonon specific heat of $Sr_{14}Cu_{24}O_{41}$ according to

$$C_p(T) = 3k_B \frac{N}{\rho V} \int x^2 \frac{\exp(x)}{(\exp(x)-1)^2} g(\varpi) d\varpi, \tag{D1}$$

where $x = \hbar\omega/k_B T$, $k_B$ is Boltzmann constant, $\omega$ is the angular frequency of phonons, $N$ is the number of atoms in a unit cell, $V$ is the volume of the unit cell, and $g(\varpi)$ is phonon DOS. For energy below 40 meV, $g(\varpi)$ was obtained from the phonon DOS measured with $E_i$ = 50 meV; while $g(\varpi)$ above 40 meV is from the phonon DOS measured with $E_i$ = 200 meV. As the phonon DOS below 3 meV are extrapolation, the calculated $C_p$ data below 40 K are less reliable than the data at higher temperatures. However, such extrapolation does not influence the calculated magnon $C_p$ near its peak at a relatively high temperature of 100 K. In addition, it is noted that the calculated phonon $C_p(T)/T^3$ does not become a constant at low temperature in the plot, which is due to limited resolution of the numerical integration with an energy step of 0.5 meV.

**Figures**

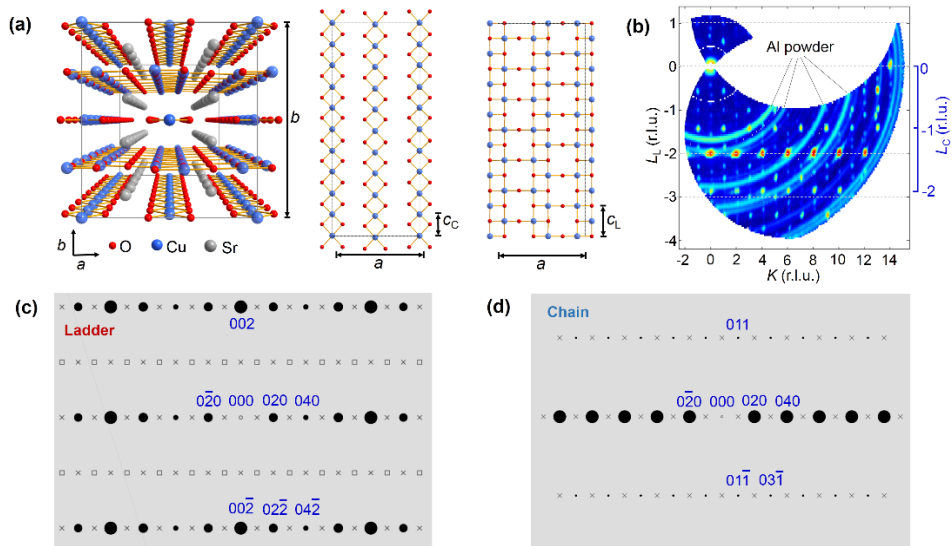

FIG. 1. (Color online) (a) Schematic illustrations of the crystal structure of $Sr_{14}Cu_{24}O_{41}$. (b) Reciprocal space map of elastic scattering in the (0$KL$) plane from ARCS measurements with the incident neutron energy $E_i$ = 30 meV at $T$ = 5 K. The $K$ and $L$ values are shown in the reciprocal lattice units (r.l.u.) for the chain and ladder sublattices indicated by the C and L subscripts, respectively. (c,d)The calculated neutron diffraction patterns for the ladder and chain sublattices in the (0$KL$) plane. Solid circles represent observable diffraction peaks. Crosses and squares are diffraction peaks with zero intensity. Crosses are zone boundaries while squares and circles are zone centers.

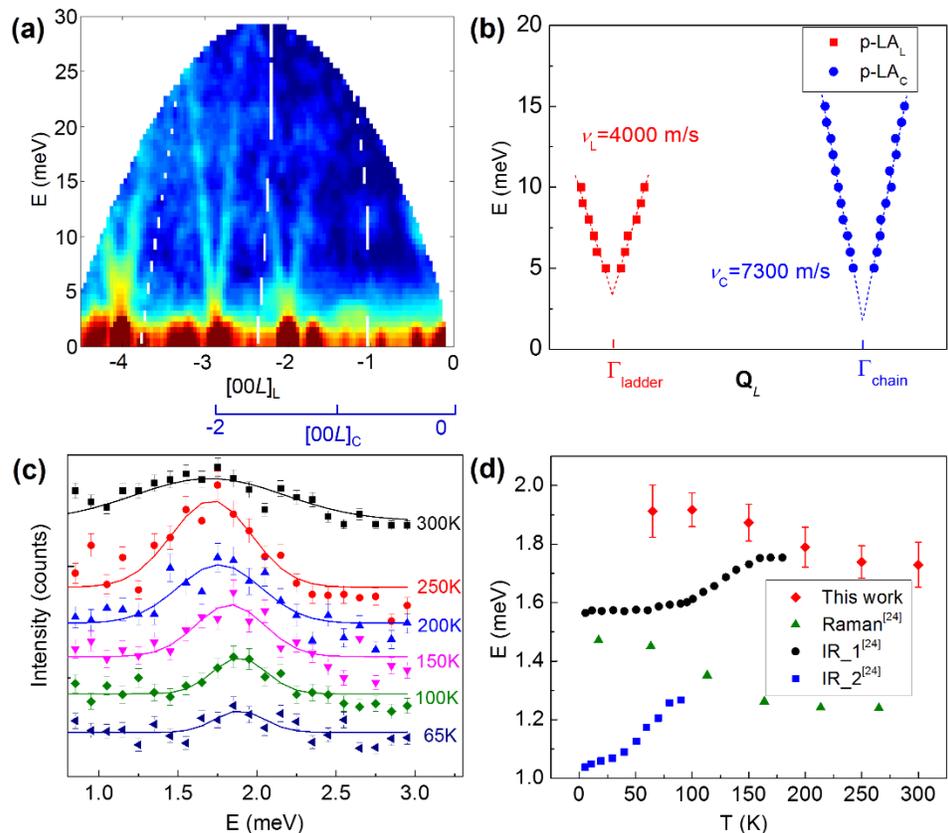

FIG. 2. (Color online) Measured INS signal of the $Sr_{14}Cu_{24}O_{41}$ single crystal showing the c-polarized acoustic-like modes. (a) $S(\mathbf{Q},E)$ data at 5K along $[00L]_L$ with $E_i = 50$ meV showing the two sets of LA-like dispersions associated with the chain and ladder sublattices. (b) Peak values of $S(\mathbf{Q},E)$ data of the two LA-like dispersions from (a). The dotted straight lines are linear fits of the measurement data. (c) Constant-$\mathbf{Q}$ INS data measured at $(011)_C$ zone center with the triple-axis spectrometer at different temperatures. (d) The Gaussian fitting of the peak position in (c) as a function of temperature. Shown for comparison are the reported Raman and IR measurement results of $Sr_{14}Cu_{24}O_{41}$.[24]

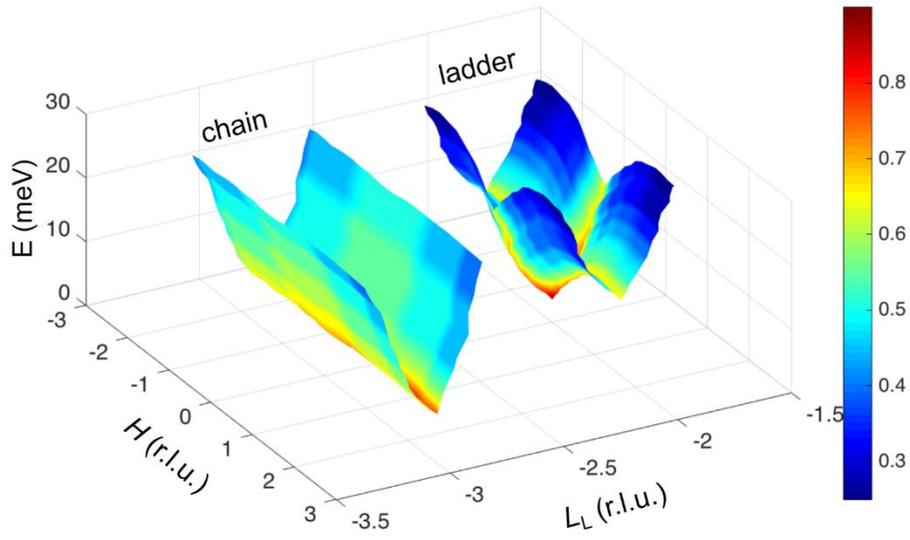

FIG. 3. (Color online) Three-dimensional (3D) plot of the phonon dispersion in the (*H*0*L*) plane for polarization parallel to the *c* axis. Portions of the dispersion surfaces for pseudoacoustic phonon polarizations are indicated.

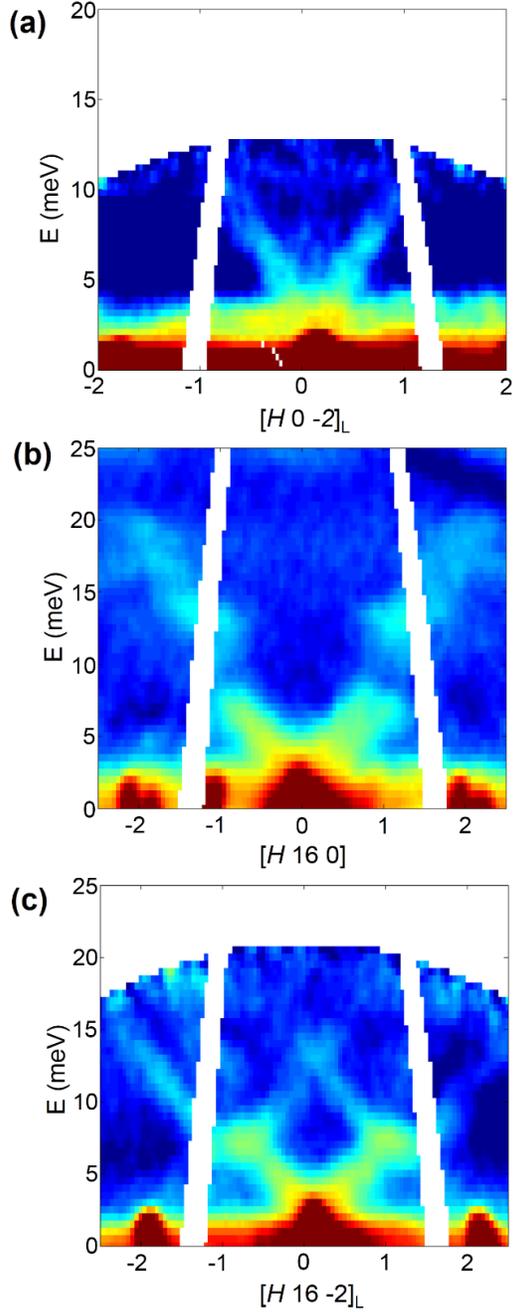

FIG. 4. (Color online) Measured INS signal of the $Sr_{14}Cu_{24}O_{41}$ single crystal for wave vectors along *a* at 5 K. (a) $S(\mathbf{Q},E)$ data along $[H\ 0\ -2]_L$ with $E_i = 30$ meV showing the *c*-polarized TA-like mode of the ladders. (b) $S(\mathbf{Q},E)$ data along $[H\ 16\ 0]_L$ with $E_i = 50$ meV showing the *b*-polarized TA mode. (c) $S(\mathbf{Q},E)$ data along $[H\ 16\ -2]_L$ with $E_i = 50$ meV showing the low-energy optical modes.

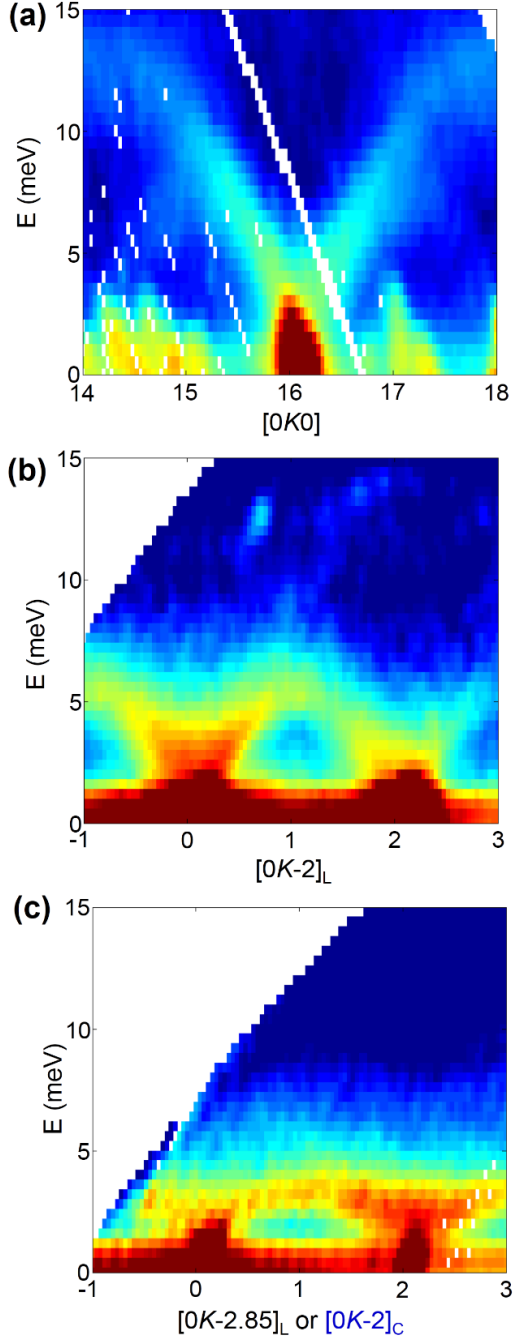

FIG. 5. (Color online) Measured INS signal of the $Sr_{14}Cu_{24}O_{41}$ single crystal for wave vectors along $b$. (a) $S(\mathbf{Q},E)$ data at 5 K along $[0K0]_L$ with $E_i = 50$ meV showing the LA mode. (b,c) $S(\mathbf{Q},E)$ data at 140 K along $[0K\text{-}2]_L$ and along $[0K\text{-}2]_C$ with $E_i = 30$ meV

showing the *c*-polarized TA-like or very low-lying optical modes associated with the ladder and chain sublattices, respectively.

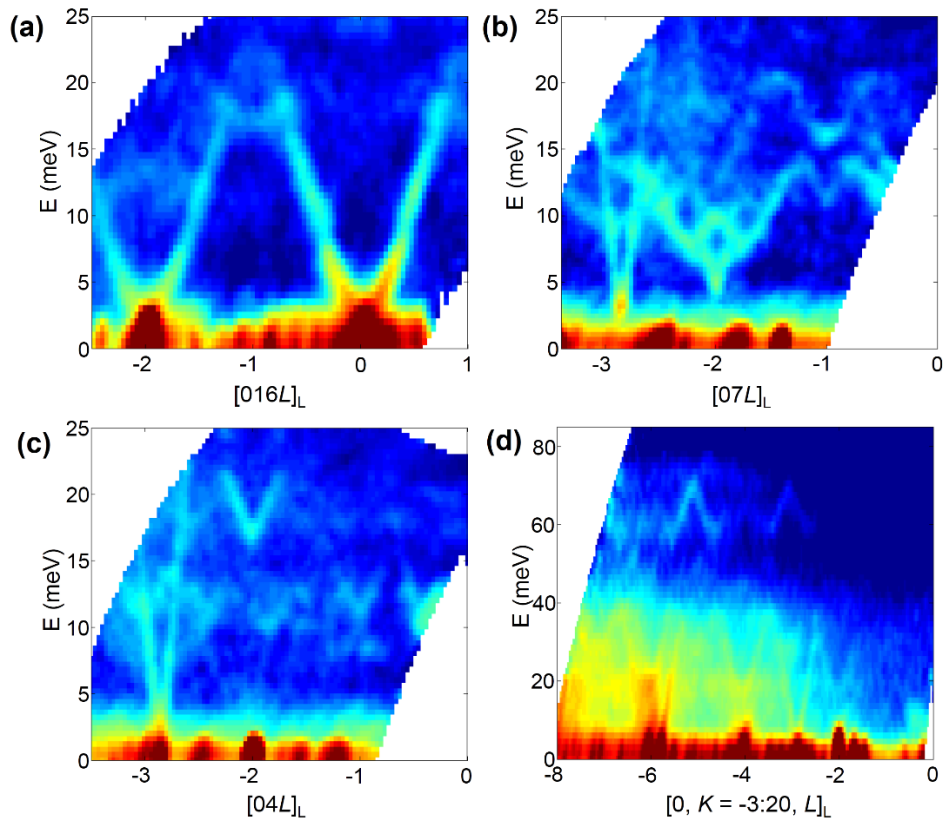

FIG. 6. (Color online) Measured INS signal of the $Sr_{14}Cu_{24}O_{41}$ single crystal along *c* at 5 K showing the TA mode and optical modes. (a) $S(\mathbf{Q},E)$ data along $[0\ 16\ L]_L$ with $E_i = 50$ meV showing the TA mode. (b,c) $S(\mathbf{Q},E)$ data along $[07L]_L$ and $[04L]_L$ with $E_i = 30$ meV showing the low-lying optical modes as well as the *c*-polarized LA-like modes of the two sublattices. (d) $S(\mathbf{Q},E)$ data obtained by integrating *K* from -3 to 20 with $E_i = 150$ meV, showing the high-energy optical modes.

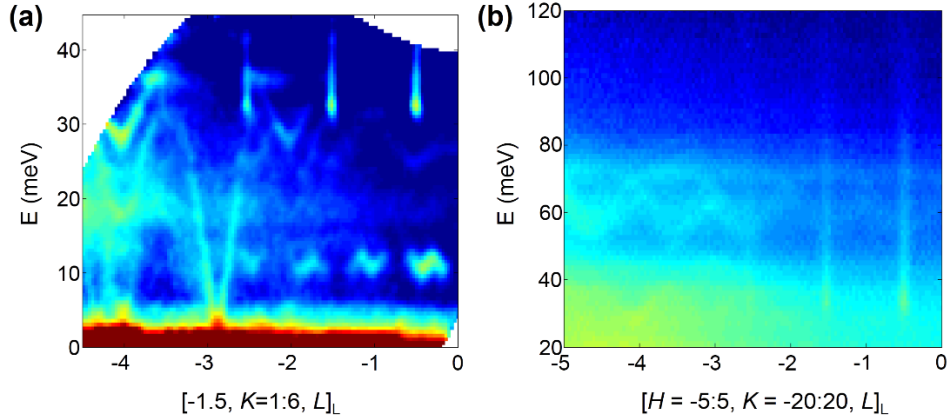

FIG. 7. (Color online) Measured INS signal of the $Sr_{14}Cu_{24}O_{41}$ single crystal at 5 K showing the magnon dispersion of the ladders and phonon branches. (a) $S(\mathbf{Q},E)$ data along $c$ with $E_i = 50$ meV integrated from 1 to 6 in $K$, showing the spin-triplet dispersions from both the ladder and chain sublattices in addition to the LA-like mode for the chains and optical modes. (b) $S(\mathbf{Q},E)$ data along $c$ with $E_i = 150$ meV integrated from -5 to 5 in $H$ and -20 to 20 in $K$ showing the spin-triplet dispersion from the ladder sublattice and high-energy optical phonons.

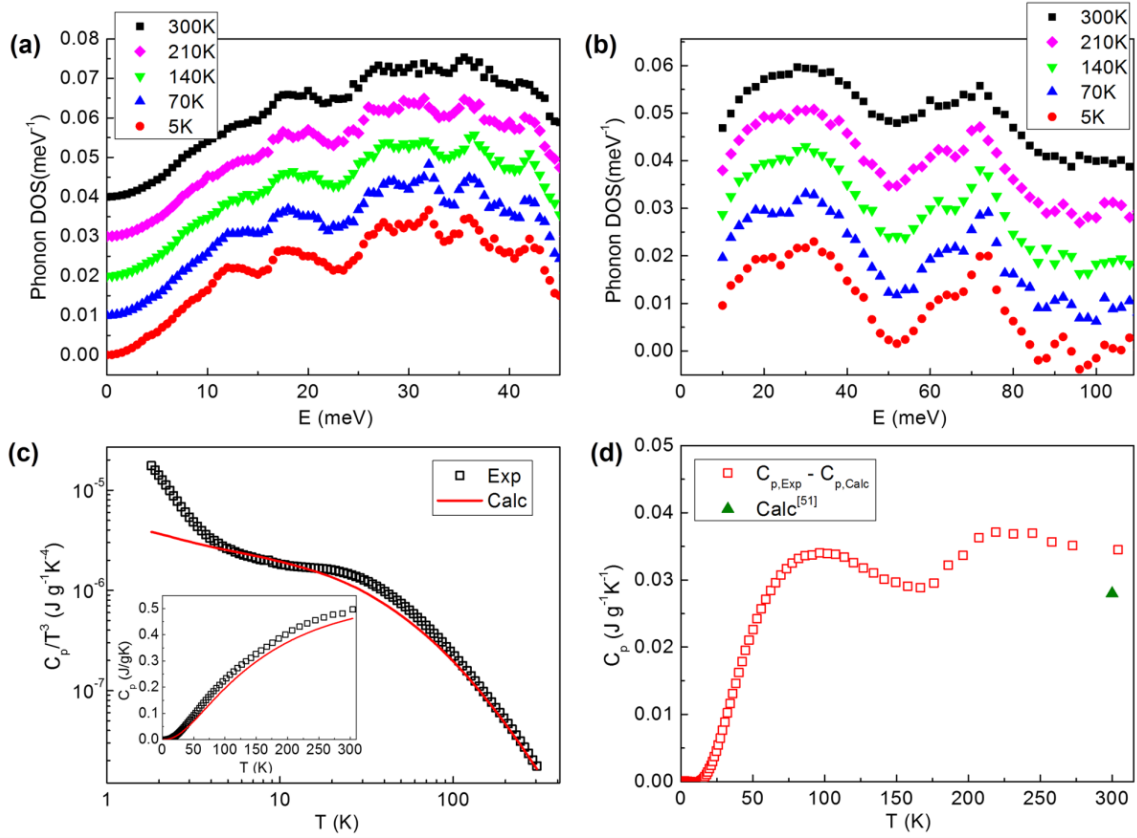

FIG. 8. (Color online) Neutron weighted phonon DOS of the $Sr_{14}Cu_{24}O_{41}$ polycrystalline sample at different temperatures for incident neutron energies of (a) 50 meV and (b) 200 meV. (c) Temperature dependence of the measured and calculated specific heat data of $Sr_{14}Cu_{24}O_{41}$ in a $C_p(T)/T^3$ versus $T$ plot. The inset of (c) is the $C_p(T)$ versus $T$ plot. (d) The difference between $C_{p,Exp}$ and $C_{p,Calc}$ in comparison with the magnon specific heat calculated by Montagnese et al.[51]

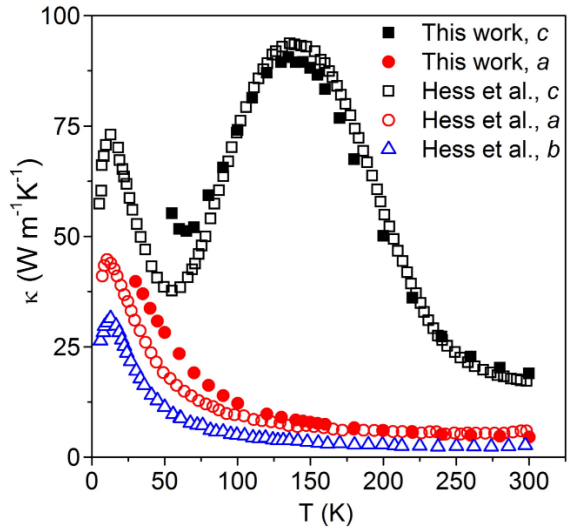

Fig. 9. (Color online) Thermal conductivity of the $Sr_{14}Cu_{24}O_{41}$ single crystal along different axes measured in this work and reported by Hess *et al.*[39] The uncertainty of the thermal conductivity measurement of the present work is 15%.

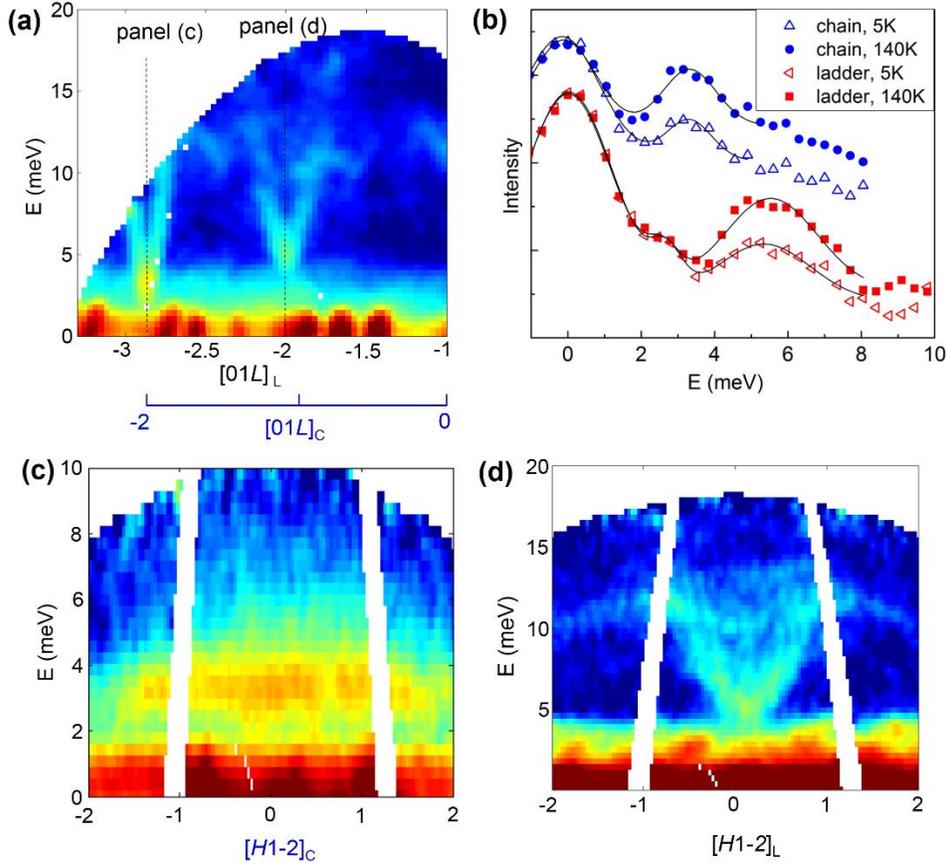

Fig. 10. (Color online) Measured INS signal of the $Sr_{14}Cu_{24}O_{41}$ single crystal at 5 K showing the *c*-polarized low-lying optical modes at the zone boundary. (a) $S(\mathbf{Q},E)$ data along $[01L]_L$ with $E_i = 30$ meV showing the energy gaps for the two optical modes. The dashed lines show the positions of the cuts in (c) and (d). (b) Energy-dependent INS intensities measured at a constant $\mathbf{Q} = (01\text{-}2)_L$ and $\mathbf{Q} = (01\text{-}2)_C$, respectively, at 5 K and 140 K. The solid lines are the peak fits of the measurement data. (c,d) $S(\mathbf{Q},E)$ data along $[H1\text{-}2]_C$ and $[H1\text{-}2]_L$, respectively, with $E_i = 30$ meV.

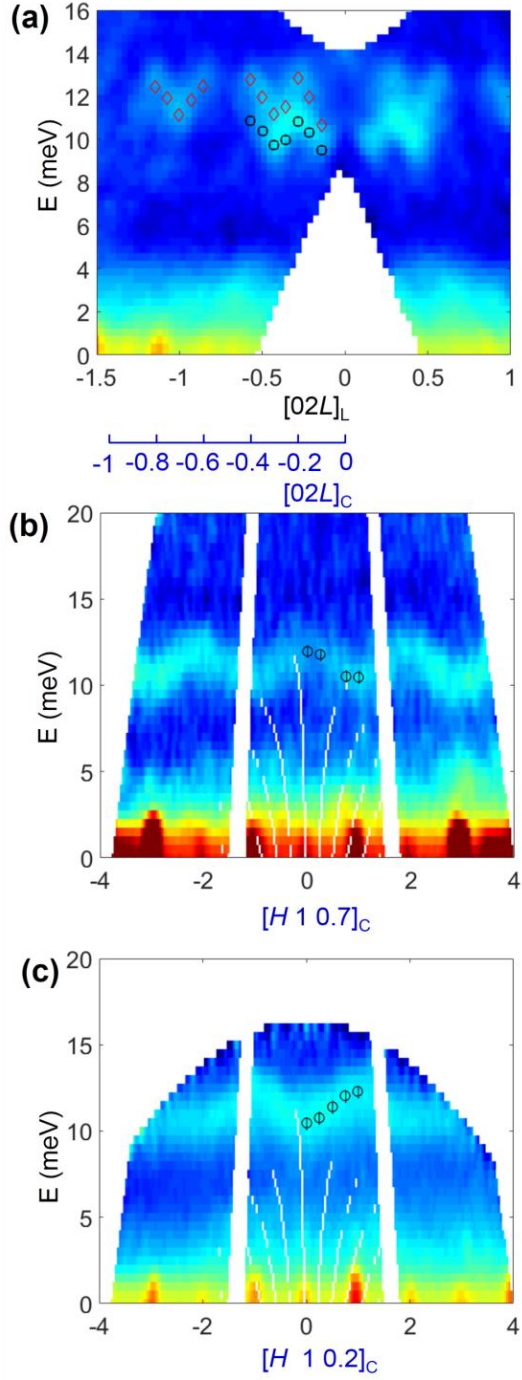

FIG. 11. (Color online) Measured INS signal of the $Sr_{14}Cu_{24}O_{41}$ single crystal at 5 K showing the magnon dispersion of the chains. (a) $S(\mathbf{Q},E)$ data along $c$ with $E_i = 30$ meV showing the low-energy magnon dispersions. The open symbols are the INS results reported by Matsuda et al.[29] (b,c) $S(\mathbf{Q},E)$ data along $a$ with $E_i = 50$ meV showing the two

low-energy magnon dispersions. The open symbols are the INS results reported by Regnault et al.[28]

**Table I.** Group velocities of the acoustic and acoustic-like phonons along different axes and polarized along the *a*, *b*, or *c* axes based on the INS data compared to the sound velocities derived from resonant ultrasound spectroscopy[42].

| Branch | INS $v_g$ (m s$^{-1}$) | $v_g$ (m s$^{-1}$) from Ref. [42] |
|---|---|---|
| **[00*L*]** | | |
| LA | - | 6430 |
| LA$_L$ | 4000 | - |
| LA$_C$ | 7300 | - |
| *b*-TA | 2100 | - |
| **[0*K*0]** | | |
| LA | 3280 | - |
| *c*-TA$_L$ | 2200 | - |
| *c*-TA$_C$ | 1500 | - |
| **[*H*00]** | | |
| LA | - | 6590 |
| *b*-TA | 3300 | - |
| *c*-TA$_L$ | 3800 | - |
| *c*-TA$_C$ | - | - |
| *c*-TA | - | 3280 |